\documentclass[superscriptaddress,twocolumn,showpacs,aps,floatfix]{revtex4}
\usepackage{graphicx}
\usepackage{natbib}
\usepackage{amsmath}
\begin{document}

\newcommand{\D}{\displaystyle} %normal formulas
\newcommand{\T}{\textstyle} %for large font
\newcommand{\SC}{\scriptstyle} %footnote
\newcommand{\SSC}{\scriptscriptstyle} %footnote to footnote

%
%
%Formeln
\def\simle{\lower 2pt \hbox {$\buildrel < \over {\scriptstyle \sim }$}}
\def\simge{\lower 2pt \hbox {$\buildrel > \over {\scriptstyle \sim }$}}
\def\intunits{{\rm s}^{-1}\,{\rm sr}^{-1} {\rm cm}^{-2}}
\def\diffunits{{\rm GeV}^{-1}\,{\rm s}^{-1}\,{\rm sr}^{-1} {\rm cm}^{-2}}
\def\epkin{\left(\frac{E_{\rm p}-m_p\,c^2}{\rm GeV}\right)}
\def\ep{E_{\rm p}}
\def\ecr{E_{\rm CR}}
\def\ensource{E_{\nu,z}}
\def\en{E_{\nu}}

%%%%%%%%%%%%%%%%%%%%%%%%%%%%%%%%%%%%%%%%%%%%%%%%%%%%%%%%%%%%%%%%
\title{High-energy neutrinos from radio galaxies}
%%%%%%%%%%%%%%%%%%%%%%%%%%%%%%%%%%%%%%%%%%%%%%%%%%%%%%%%%%%%%%%%
\author{J.~Becker Tjus} \vspace*{-20pt}
\affiliation{Theoretische Physik IV: Plasma-Astroteilchenphysik, Fakult\"at f\"ur Physik \& Astronomie,
  Ruhr-Universit\"at Bochum, Germany}
\author{B.\ Eichmann} \vspace*{-20pt}
\affiliation{Theoretische Physik IV: Plasma-Astroteilchenphysik, Fakult\"at f\"ur Physik \& Astronomie,
  Ruhr-Universit\"at Bochum, Germany}
\author{F.\ Halzen} \vspace*{-20pt}
\affiliation{Department of Physics, University of Wisconsin, Madison, WI 53706, USA}
\author{A.\ Kheirandish} \vspace*{-20pt}
\affiliation{Department of Physics, University of Wisconsin, Madison, WI 53706, USA}
\author{S.~M.~Saba} \vspace*{-20pt}
\affiliation{Theoretische Physik IV: Plasma-Astroteilchenphysik, Fakult\"at f\"ur Physik \& Astronomie,
  Ruhr-Universit\"at Bochum, Germany}
\begin{abstract}
The IceCube experiment has recently reported the first observation of
high-energy cosmic neutrinos. Their origin is still unknown. In this
paper, we investigate the possibility that they originate in active
galaxies. We show that hadronic interactions (pp) in the generally less powerful, more frequent, FR-I radio galaxies are one of the candidate source classes being able to accommodate the observation while the more powerful, less frequent, class of FR-II radio galaxies
has too low of a column depths to explain the signal.
\end{abstract}
%PACS used:
%  98.54.Gr 	Radio galaxies
% 98.70.Sa Cosmic rays (including sources, origin, acceleration, and interactions)
% 95.55.Vj 	Neutrino, muon, pion, and other elementary particle detectors; cosmic ray detectors (see also 29.40.-n Radiation detectors—in Nuclear physics)
% 95.85.Ry 	Neutrino, muon, pion, and other elementary particles; cosmic rays
% 98.54.Cm 	Active and peculiar galaxies and related systems (including BL Lacertae objects, blazars, Seyfert galaxies, Markarian galaxies, and active galactic nuclei)
\maketitle
%%%%%%%%%%%%%%%%%%%%%%%%%%%%%%%%%%%%%%%%
\section{Introduction}
%%%%%%%%%%%%%%%%%%%%%%%%%%%%%%%%%%%%%%%%
The search for the origin of cosmic rays began with their first detection in 1912. Recently, a first step toward the identification of the cosmic ray sources has been done by a first evidence of a high-energy neutrino signal in the IceCube detector \cite{icecube_evidence2013}. The measurement of 28 events in a search using a veto for particles (muons and neutrinos) produced in atmospheric air showers provides a significance of $\sim 4\,\sigma$ and corresponds to an astrophysical flux of $E^2\,dN/dE =(1.2\pm0.4)\cdot 10^{-8}\,\diffunits$ \cite{icecube_evidence2013}. At this early stage, the directional information does not suffice to clearly identify the sources. Different implications from the detection of astrophysical high-energy neutrinos for various emission models have been discussed most recently, for instance concerning the emission of neutrinos from gamma-ray bursts \cite{cholis2013,winter2013}, during extragalactic propagation \cite{kalashev2013,roulet2013}, photohadronic interaction in AGN \cite{winter2013,stecker2013}, as well as proton-proton interactions in galaxy clusters and starburst galaxies/ULIRGs \cite{he2013,murase2013}. Sources of galactic emission are considered in \cite{anchordoqui2013,gonzalez2013,razzaque2013}.

In this paper, we investigate the hypothesis of active galactic nuclei (AGN) being the sources of ultra-high energy cosmic rays (UHECRs). 
%If these extragalactic sources give rise to the observed flux of astrophysical high-energy neutrinos, the environment for proton-proton interactions is well-%defined within an order of magnitude. 
AGN have long been discussed as one of the few possible source classes
being able to accelerate particles up to the observed maximum energies
of around $10^{20}-10^{21}$~eV \cite{biermann_strittmatter1987}. There exist different acceleration scenarios and the unified AGN model allows for different sub-AGN/classes to possibly be the dominant source of UHECRs. Both intrinsic properties as well as the orientation of the objects play a role in this respect. For a summary of a discussion concerning AGN sub-classes as neutrino emitters, see \cite{becker2008}. In particular, radio loud AGN are typically discussed as interesting candidates: although these only make up a fraction of about 10\% of the entire AGN population, they have very powerful radio jets, not provided by radio quiet galaxies like Seyferts. Among radio loud galaxies, FR-I and FR-II type AGN are among the most prominent candidates, having powerful radio jet and being very frequent among the radio loud class of AGN.

A first hint of an anisotropy in the cosmic ray distribution at Earth was announced in \cite{auger2007}, where UHECRs above $6\cdot 10^{19}$~eV appear to show some correlation with the distribution of local AGN (within a distance of $\sim 75$~Mpc): as the flux of UHECRs at larger distances is expected to be absorbed at those energies by interactions with the Cosmic Microwave Background (CMB), such a clustering would be expected if AGN are the sources of UHECRs. Although there has not been a clear confirmation of the signal yet, the anisotropy persists at a low level and the nearest AGN Centaurus A - an FR-I type AGN - is discussed to be responsible for a large fraction of the correlated events \cite{giacinti2011,anchordoqui_cena2011,biermann_cena_2012}. The detection of high-energy gamma-rays from Centaurus A \cite{hess_cena2009,rieger2009,rieger2012} could be another hint for pion production in AGN, see e.g.\ \cite{saba2013}, but it is not yet confirmed if the origin of the gamma-rays is of hadronic or leptonic nature. Neutrinos, on the other hand, must be of hadronic origin and high-energy neutrino detection therefore provides a unique method to identify the sources of UHECRs. 

Cosmic rays have been discussed to be able to be accelerated at different sites in AGN. Their acceleration in AGN cores would lead to photohadronic production of neutrinos \cite{stecker1991,nellen1993}. 
Shock acceleration in knots of AGN jets as they are observed in FR-I galaxies, or in the termination shock of the jet with the intergalactic medium as seen in FR-II galaxies, are have been discussed as possible cosmic ray acceleration sites, see e.g.\ \cite{biermann_strittmatter1987,bbr2005,meli2007,becker_biermann2009,eichmann2012}. These sites are connected to a specific column depth, and so, cosmic ray interactions with matter are a necessary consequence of each acceleration scenario. The central question is what intensity to expect from which acceleration site. Thus,
with the very first evidence of a high-energy neutrino signal, even
without the knowledge of any directional information, neutrino
astronomy can now be used to provide constraints on the emission
region within AGN, as we will show at the example of radio galaxies in
this paper. We discuss the straight-forward way of how to calculate
the column density of the emission region by combining information from the measured flux, the ratio between electron- and proton-luminosity, $f_e$, and the distribution of AGN in the Universe. 

In this paper, we focus on radio galaxies. Another option would be to discuss blazars, where you look at the boosted emission of the jet by directly looking into it. A model of blazar emission is presented in \cite{murase2014}. The focus here lies on the modeling of the photon fields and has difficulties to explain the IceCube results. We refrain from modeling these blazars as
well as the effects of photohadronic emission in order to keep ourselves to as little parameters as possible: as for the blazars, both the luminosity function and boosting effects lead to relatively high uncertainties. Concerning photohadronic emission scenarios, a primary source of uncertainty comes from the composition
of cosmic rays. For a large fraction of heavy nuclei, the neutrino flux is significantly reduced with respect to a pure proton flux. In addition, the spectral shape of the neutrino spectrum from photohadronic interactions is highly sensitive to the shape and bandwidth of the target photon field. The main effect comes from the fact
that a delta resonance needs to be produced. This is a threshold effect depending on the bandwidth of the magnetic field. For moderate boost factors, relatively high-energy photon fields are needed in order to lower the spectral break to below the IceCube bandwidth of the detected signal: The IceCube observed bandwidth reaches up to $\sim 1$~PeV neutrino energy, which corresponds to a proton energy of $\sim 20$~PeV. Thus, in order to have a flat spectrum, i.e.\ close to $E^{-2}$, a photon field with a significant contribution at an energy $\epsilon_{\gamma}$ needs to be present with the condition
\begin{equation}
E_{p}\cdot \epsilon_{\gamma}=\frac{m_{\Delta}^{2}-m_{\rm p}^{2}}{4}\cdot \frac{\Gamma}{(1+z)}
\end{equation}
Here, the proton energy is given at Earth in the observer's frame, and is therefore boosted with the Lorentz factor of the production region, in AGN at least $\Gamma\sim 10$, or higher. It is also corrected for redshift. Considering that the break needs to be at $E_{\rm p}<20$~PeV in order to have a spectrum close to $E^{-2}$ in the IceCube bandwidth, the photon field at the source needs to have a significant contribution at above $40\cdot (\Gamma/10)\cdot (2/(1+z))$~eV. Here, conservative values for the average boost factor and redshift have been assumed. It is therefore very difficult to receive an $E^{-2}-$type neutrino spectrum from photohadronic interactions in the relevant energy range.
Therefore, we only consider radio galaxies and proton-proton interactions here. Of course, photohadronic emission could potentially lead to an additional contribution to the neutrino flux at higher energies and in our conclusions, we discuss what this means for our predictions.

%%%%%%%%%%%%%%%%%%%%%%%%%%%%%%%%%%%%%%%%
\section{The neutrino flux at the source}
%%%%%%%%%%%%%%%%%%%%%%%%%%%%%%%%%%%%%%%%
Pions are produced in proton-proton interactions via $p\,p\rightarrow
\pi^{0/\pm}$ and neutrinos are produced subsequently via the decay of
the charged pions, see
    e.g.\ \cite{becker2008} for a review. In the following
    calculation, we formally follow the paper of
    \cite{mannheim_schlickeiser1994}. It should be noted that monte
    carlo approaches like SIBYLL, QGSJet, EPOS or DPMJet provide much
    more detailed and up-to-date particle physics. However, the
    uncertainty included by using the analytic approximation is
    rather small when compared to the astrophysical uncertainties
    discussed in this paper. Therefore, we use the delta-functional
    approach here. In the approach sketched by
\cite{mannheim_schlickeiser1994}, the cross section for proton-proton interactions is assumed to be constant, $\sigma_{\rm pp}\approx 3\cdot 10^{-26}$~cm$^{2}$ and the pion production efficiency of protons with an energy $E_p$ that is above the threshold energy $E_{\rm th}$ is given as
\begin{equation}
\xi_{\pi^{\pm}}=2\cdot \left(\frac{E_{\rm p}-E_{\rm th}}{\rm GeV}\right)^{1/4}\,.
\end{equation} 
Consequently, the number of pions per energy and time interval $q_{\pi^{\pm}}(E_\pi)$ is related to the proton rate $q_{\rm p}(E_{\rm p},\,\tau)$ according to 
\begin{equation}
q_{\pi^{\pm}}=\int_{E_{\rm th}}^{\infty}dE_{\rm p}\,\xi_{\pi^{\pm}}\,\delta\left(E_{\pi}-\left<E_{\pi}\right>\right)\int_0^{\tau}d\tau'\,q_{\rm p}(\tau')\,,
\end{equation}
 where it is assumed that all energy is going to the average pion, $E_{\pi}\approx \left<E_{\pi}\right>$.
Due to pion production the proton rate is determined by the optical depth $\tau$, which yields $q_{\rm p}(\ep,\tau)=j_{\rm p}(\ep)\,\exp(-\tau)$ with the undamped rate $j_{\rm p}(\ep)$, so that the pion rate {\it at the source} is described as:
\begin{equation}
 q_{\pi^{\pm}}= \int_{E_{\rm th}}^{\infty}dE_{\rm p}
 \left(1-\exp(-\tau)\right)  j_{\rm p}(E_{\rm p})\,\xi_{\pi^{\pm}}\,\delta\left(E_{\pi}-\left<E_{\pi}\right>\right)\,.
\end{equation}
Approximating for low optical depths, $\tau=l\cdot n \cdot \sigma_{\rm pp} <1$, it follows
\begin{equation}
 q_{\pi^{\pm}}= 1.6\cdot n_H\cdot l \cdot \sigma_{\rm pp}\cdot \int_{E_{\rm th}}^{\infty}d\ep\,\,  j_{\rm p}\, \xi_{\pi^{\pm}}\,\delta\left(E_{\pi}-\left<E_{\pi}\right>\right)\,.
\end{equation}
using $n\approx 1.6\,n_H$, which takes H-I, H-II and H$_2$ as well as
He into account \cite{mannheim_schlickeiser1994}. While we assume only
protons here, the result is not subject to change due to different
composition scenarios, due to a simple scaling of the cross section
with the mass number of the particle as discussed in
\cite{dermer1986} and references therein. Here, $l$ is the length scale the cosmic rays traverse through the dense medium. The product of the density and the length scale can be abbreviated as the column density, $N_H=l\cdot n_H$.
The threshold energy is close to the proton mass and we approximate $E_{\rm th}\approx m_{\rm p}\,c^2$. 

The differential proton number per energy and time interval at the source is given as
\begin{equation}
j_{\rm p}(\ep)=A_{\rm p}\cdot \left(\frac{\ep-m_p\cdot c^2}{{\rm GeV}}\right)^{-p}\,.
\end{equation}
Substituting $x:=\left<E_{\pi}\right>=\frac{1}{6}\epkin^{3/4}\,{\rm GeV}$ gives a pion spectrum at the source of
\begin{equation}
q_{\pi^{\pm}}(E_\pi)\approx 26\cdot N_H\cdot A_{\rm p}\cdot \sigma_{\rm pp}\cdot \left(\frac{6\cdot E_{\pi}}{\rm GeV}\right)^{-\frac{4}{3}(p-\frac{1}{2})}\,.
\end{equation}
The total neutrino rate {\it at the source} is then given by the sum of the first muon neutrino (directly from the pion), the second muon neutrino and the electron neutrino, both from the muon decay, 
\begin{equation}
q_{\nu,{\rm tot}}=q_{\nu_{\mu}}^{(1)}+q_{\nu_{\mu}}^{(2)}+q_{\nu_{e}}\,.
\end{equation}
The neutrino spectra are received from the pion spectrum by assuming
that the total energy of the pions is distributed equally among the
four produced particles \cite[e.g.]{kelner2006},
\begin{equation}
q_{\nu_{i}}(E_{\nu_{i}})=q_{\pi}(4\,E_{\nu_{\i}})dE_{\pi}/dE_{\nu_{i}}=4\cdot
q_{\pi}(4\ E_{\nu_{i}})
\end{equation}
for each neutrino,
$\nu_{i}=\overline{\nu}_{e}/\nu_{e},\,\nu_{\mu},\,
\overline\nu_{\mu}$. Here, it depends on the charge of the pion if an
electron or an anti-electron neutrino is produced. As IceCube does not
distinguish between neutrinos and anti-neutrinos, we will neglect this
piece of information in the following. 

In the above calculation of the neutrino rate from the pion rate, we assumed that the
number of a single neutrino flavor produced in the infinitesimally
small energy bin
$dE_{\nu}$ arises from the original pion in the energy bin
$dE_{\pi}=4\cdot dE_{\nu}$ and use that one neutrino of a fixed flavor
is produced in the final state of the pion decay: $q_{\nu}(E_{\nu})\,dE_{\nu}=q_{\pi}(4\,E_{\nu})dE_{\pi}$. %q_{\nu_{\mu}}^{(1)}\simeq q_{\nu_{\mu}}^{(2)}\simeq q_{\nu_{e}}\simeq \frac{1}{4}\cdot q_{\pi^{\pm}}(E_{\nu})$.

The total neutrino rate at the source becomes
\begin{equation}
q_{\nu,{\rm tot}}\approx 3\cdot 10^{2}\cdot N_H\cdot A_{\rm p}\cdot \sigma_{\rm pp}\cdot \left(\frac{24\cdot \en}{{\rm GeV}}\right)^{-\frac{4}{3}p+\frac{2}{3}}\,.
\label{nusource}
\end{equation}
This simple analytic result represents an approximation of the
particle-physical processes with only smaller deviations when using
the full representation of energy-dependent cross section and full
energy distribution for one interaction. The effects are described in
\cite{kelner2006}, and could easily be taken into account here. We
consciously chose not to do this however, for a simple reason: when
using the semi-analytically approximation of \cite{kelner2006}, the
spectral shape does not represent a pure power-law anymore, but
deviates somewhat from the original behavior. This is an effect of the
moderately increasing cross section and the energy distribution for a
single interaction. Thus, the comparison of our results to the flux
estimate from IceCube, which is given as a pure $E^{-2}-$flux becomes
much more subtle. This is not only an effect of the theory, but we
would also have to change the flux estimate accordingly: putting the
correct normalization for a spectrum with an arbitrary spectral
behavior is not straight forward and should in our opinion remain
subject of the experimental results. Therefore, in order to have a
correct comparison, we work with the delta-approximation, which
provides the spectral behavior for which the normalization is
given. This has the second advantage of easily being able to interpret
this fully analytically result.

Equation (\ref{nusource}) now provides us with an estimate of the
total neutrino flux at the source. The spectral behavior of the
protons can be estimated from diffusive shock acceleration and we take
it to be $p=2$ here in order to compare with the IceCube results. The
cross section itself is a particle-physical property which is well
known compared to the astrophysical uncertainties. The column density
is one of the main free parameters in this calculation which we will
discuss later. The proton normalization is the second (relatively)
free parameter. For a single radio galaxy, it can be estimated from
radio observations as we show in the following paragraph.

%===============================================
%\section{CR normalization}
%===============================================
The normalization of the cosmic ray spectrum can be estimated from the following considerations: The radio luminosity of AGN, $L$, provides a measure for the AGN luminosity in electrons. The electron luminosity is equal to or larger than the radio luminosity of the source, as the latter is produced when electrons are accelerated and emit synchrotron radiation: $L_{\rm e}=\chi\cdot L$ with $\chi \geq 1$. 

Hadronic cosmic rays and electrons are connected via a constant fraction $f_e$, $L_{\rm e}=f_e\cdot L_{\rm p}$,
\begin{equation}
L_{\rm p}=\int j_{\rm p}(\ep) \ep\,d\ep\approx \frac{\chi\cdot L}{f_e}\,.
\end{equation}
 The normalization of the CR spectrum is therefore:
\begin{equation}
A_{\rm p}=A_{\rm p}(L,z)= \frac{\chi}{f_e}\cdot \left[\ln\left(E_{\max}/E_{\min}\right)\right]^{-1}\cdot  L\,{\rm GeV}^{-2} 
\label{norm}
\end{equation}
For the case of $p\neq2$,
\begin{eqnarray}
A_{\rm p}&=&A_{\rm p}(L,z)= \frac{\chi}{f_e}\cdot\frac{1}{-p+2}
\label{norm_neq_2}
\\
&\cdot& \left[\left(\frac{E_{\max}}{\rm
    GeV}\right)^{-p+2}-\left(\frac{E_{\min}}{\rm GeV}\right)^{-p+2}\right]^{-1}\nonumber\\
&\cdot&  L\cdot {\rm GeV}^{-2} \nonumber
\end{eqnarray} 
The uncertainties in the parameters of this result will be discussed later in this paper.
%%%%%%%%%%%%%%%%%%%%%%%%%%%%%%%%%%%%%%%
\section{The diffuse neutrino flux from AGN}
%%%%%%%%%%%%%%%%%%%%%%%%%%%%%%%%%%%%%%%

The {\it diffuse neutrino flux at Earth} is given as
\begin{equation}
\Phi_\nu=\int_{L}\int_{z}\frac{q_{\nu,tot}}{4\,\pi\,d_L(z)^{2}}\cdot \frac{dn_{\rm AGN}}{dV\,dL}\cdot \frac{dV}{dz}\,dz\,dL\,.
\label{diffuse}
\end{equation}
Here, $d_L$ is the luminosity distance, d$n_{\rm AGN}$/(d$V\,$d$L$) is the radio luminosity function of the AGN and d$V$/d$z$ is the comoving volume at a fixed redshift $z$. The radio luminosity function is usually represented by the product of a luminosity-dependent and a redshift-dependent function, dn$_{\rm AGN}/(d$V$\,d$L)$=g(L)\cdot f(z)$. Including the single source flux (Equ.\ (\ref{nusource})) and the representation for the cosmic ray spectrum normalization given in Equ.\ (\ref{norm}), the diffuse neutrino flux can be parametrized as
\begin{equation}
\Phi_\nu=\zeta_c\cdot \zeta_z\cdot \zeta_L\cdot
\left(\frac{E_{\nu,0}}{\rm GeV}\right)^{-\frac{4}{3}p+\frac{2}{3}}\,.
\label{prediction}
\end{equation}
Here, we used the adiabatic energy losses, $E_{\nu}=(1+z)\cdot E_{\nu,0}$, with $E_{\nu}$ as the energy at the source and $E_{\nu,0}$ the energy at the detector.
The above introduced factors represent:
\begin{eqnarray}
\zeta_{c}
&\approx&2.4\cdot
  10^{-4}\cdot24^{-\frac{4}{3} p+\frac{2}{3}}\,{\rm
  GeV}^{-2}\nonumber\\
&\cdot&\left\{\begin{array}{lll}
\frac{1}{-p+2}\cdot \left[\left(\frac{E_{\max}}{\rm
    GeV}\right)^{-p+2}-\left(\frac{E_{\min}}{\rm
    GeV}\right)^{-p+2}\right]^{-1}&&{\rm for }\, p\neq 2\\
\ln\left[\frac{E_{\max}}{E_{\min}}\right]^{-1}&&{\rm for }\, p=2\end{array}\right.\nonumber\\
&\cdot& \left(\frac{\chi}{f_e}\right)\cdot \left(\frac{N_H}{10^{20}{\rm cm}^{-2}}\right)\\
\zeta_{L}&=&\int_{L_{\min}}^{L_{\max}}g(L)\cdot L\,dL\\
\zeta_{z}&=&\int_{z_{\min}}^{z_{\max}}\frac{1}{4\,\pi\,d_{L}^{2}\cdot (1+z)^{\frac{4}{3}p-\frac{2}{3}}}\cdot f(z)\,\frac{dV}{dz}\,dz\,.
\end{eqnarray}
Above, it is assumed that the energy range is
$\ln\left(E_{\max}/E_{\min}\right)\approx 6$, assuming approximately 3
orders of magnitude between minimal and maximal energy. This range
corresponds to the observed UHECR spectrum and probably extends
towards lower minimal energies, but as the behavior is logarithmic,
the expected changes are rather small and are neglected here.
It should be noted that the neutrino rate from one single source is transformed into a flux at Earth by dividing by $1/(4\,\pi\,d_{L}^{2})$ as we derive the flux from a radio luminosity given at the source. Hence, no additional redshift factor, but the redshift-dependent luminosity distance is needed, as this distance measure is defined to transform from luminosities at the source and fluxes at Earth.
%===============================================
\subsection{Radio Luminosity Function}
%===============================================

The radio luminosity function is usually expressed as the product of a redshift dependent part, $f(z)$ and a luminosity dependent part, $g(L)$,
\begin{equation}
\frac{dn_{\rm AGN}}{dV\,dL}=f(z)\cdot g(L)\,.
\label{rlf}
\end{equation}
Depending on what sub-class of AGN is considered, the behavior of the radio luminosity function can vary. In this paper, we focus on FR-I and FR-II galaxies.

Concerning FR-I and FR-II galaxies, Willott et al. (2001) \cite{willott2001} provide luminosity functions for $\left(\Omega_{\rm M},\,\Omega_{\Lambda}\right)=\left(1,\,0\right)$ and $\left(\Omega_{\rm M},\,\Omega_{\Lambda}\right)=\left(0,\,0\right)$. As the authors argue that their results for $\left(\Omega_{\rm M},\,\Omega_{\Lambda}\right)=\left(0,\,0\right)$ even reproduce a $\Lambda$CDM cosmology with $\left(\Omega_{\rm M},\,\Omega_{\Lambda}\right)=\left(0.3,\,0.7\right)$, we use their results for the $\left(\Omega_{\rm M},\,\Omega_{\Lambda}\right)=\left(0,\,0\right)$ cosmology, model C in the paper. For other redshift-dependent factors entering the calculation, we use a $\Lambda$CDM cosmology with $h=0.7$ and $\left(\Omega_{\rm M},\,\Omega_{\Lambda}\right)=\left(0.3,\,0.7\right)$.

The reference luminosity given at 0.151GHz and per steradian by
Willott et al (2001) \cite{willott2001} is converted into a total luminosity by multiplying with the frequency, $0.151$~GHz and integrating over $4\pi$.
%--------------------------------------------
\subsubsection{FR-I galaxies}
%--------------------------------------------
Generally, the luminosity-dependent part behaves as

\begin{equation}
g(L)=\frac{1}{\ln(10)\,L}\,\rho_{0}\cdot \left(\frac{L}{L_{\star}}\right)^{-\alpha}\cdot \exp\left[-(\frac{L}{L_{\star}})^{\beta}\right]
\end{equation}

Parameters for FR-I galaxies are $\rho_{0, FR-I}=10^{-7.523}$ Mpc$^{-3}$ $\Delta\log(L_{151})$, $\alpha=0.586$, $L_{\star,FR-I}=10^{42.76}$~erg/s and $\beta=1$. 
FR-II galaxies have the parameter setting $\rho_{0,FR-II}=10^{-6.757}$
Mpc$^{-3}$ $\Delta\log(L_{151})$, $\alpha_{FR-II}=2.42$ and
$L_{\star,FR-II}=10^{43.67}$~erg/s and $\beta=-1$.

The redshift dependence is parametrized as
\begin{equation}
f_{\rm FR-I}(z)=\left\{\begin{array}{lll}(1+z)^\gamma&{\rm for}&z<z_{0,FR-I}\\
(1+z_{0,FR-I})^\gamma&{\rm for}&z\geq z_{0,FR-I}\\
\end{array}\right.
\end{equation}
with $z_{0,FR-I}=0.710$ and $\gamma=3.48$. 

For FR-I galaxies, we therefore find
\begin{eqnarray}
\zeta_{L,FR-I}&=&7.8\cdot 10^{37}\,{\rm GeV/(s}\,{\rm Mpc}^3)\\
\zeta_{z,FR-I}&=&240\,{\rm Mpc}/{\rm sr}\\
\end{eqnarray}

The redshift-integrated factor $\zeta_{z,FR-I}$ (and later also $\zeta_{z,FR-II}$) is calculated in a
$\Lambda$CDM cosmology, $(\Omega_m,\,\Omega_{\Lambda})=(0.3,\,0.7)$
with $h=0.7$.
%----------------------------------------------
\subsubsection{FR-II galaxies}
%----------------------------------------------
Willot et al (2001) provide the radio luminosity function for FR-II galaxies similarly to FR-I galaxies, but with other parameters,
\begin{equation}
g_{\rm FR-II}(L))=\frac{1}{\ln(10)\,L}\,\rho_{0}\cdot \left(\frac{L}{L_{\star}}\right)^{-\alpha}\cdot \exp\left[-\frac{L}{L_{\star}}\right]\,.
\end{equation}
Most importantly, the luminosity power-law dependence behaves
inversely for the two samples. While FR-I galaxies become more
frequent towards lower luminosities, the FR-II radio luminosity
function cuts off at $L_{\star}$ and has a dominant contribution
towards high-luminosity sources. This behavior reflects the division
of FR-I and FR-II galaxies by their luminosities, FR-II galaxies
representing the high-luminosity sample with dominant emission from
the lobes, FR-I galaxies representing the low-energy sample with the
main emission along the central part of the jet.

The redshift dependence for FR-II galaxies is given as
\begin{equation}
f_{\rm FR-II}(z)=\left\{\begin{array}{lll}\exp\left(-\frac{1}{2}\left[\frac{z-z_{0,FR-II}}{z_{1}}\right]^2\right)&{\rm for}&z<z_{0,FR-II}\\
\exp\left(-\frac{1}{2}\left[\frac{z-z_{0,FR-II}}{z_{2}}\right]^2\right)&{\rm for}&z\geq z_{0,FR-II}\\
\end{array}\right.
\end{equation}
Parameters for the redshift dependence of FR-II galaxies are
$z_{0,FR-II}=2.03$, $z_{1,FR-II}=0.568$ and $z_{2,FR-II}=0.956$.

For FR-II galaxies, we find
\begin{eqnarray}
\zeta_{L,FR-II}&=&1.6\cdot 10^{39}\,{\rm GeV/(s}\,{\rm Mpc}^3)\\
\zeta_{z,FR-II}&=&4\,{\rm Mpc}/{\rm sr}\,.
\end{eqnarray}
%=======================================================
\subsection{Doppler Boosting}
%=======================================================

Effects due to possible Doppler boosting cancel out in this calculation: the radio luminosity we use in order to determine the proton density of the source is measured in the observer's frame. Thus, the additional factor based on the transformation of the luminosity from the observer's frame to the frame of the source vanishes due to the inverse transformation of the neutrino flux from the source to the observer's frame. Effects of area transformation cancel out as well, as we transform the radio luminosity per steradian to a luminosity by multiplying by an opening angle of $4\pi$ and then divide by the same factor to account for the fraction of neutrinos that reaches Earth. Both factors scale with the boost factor in the same way.

%===============================================
\section{Constraints on cosmic ray acceleration regions}
%===============================================

If we now consider a certain class of AGN as responsible for the IceCube excess, the total neutrino flux per flavor must match the observed flux,
\begin{eqnarray}
\frac{1}{3}\left(\frac{E_{\nu,0}}{\rm GeV}\right)^{2}\Phi_\nu&=&1.2\cdot 10^{-8}\,{\rm GeV}^{-1}\,{\rm cm}^{-2}\,{\rm s}^{-1}\,{\rm sr}^{-1}\\
&=&\frac{1}{3}\cdot \zeta_c\cdot \zeta_L\cdot \zeta_z\,.
\label{excess}
\end{eqnarray}
Here, the measured flux is given {\it per flavor}, which is why the results derived above need to be divided by the number of flavors.
Comparing Equ.\ (\ref{excess}) with the prediction from Equ.\ (\ref{prediction}), the column density of the interaction region in this scenario is constrained to
\begin{eqnarray}
N_{H,FR-I}\approx 10^{24.57\pm1.0}\,\left({f_e\over 0.06}\right)\,\left({100\over \chi}\right)\, {\rm cm}^{-2}\\
N_{H,FR-II}\approx 10^{25.03\pm 1.0}\,\left({f_e\over 0.06}\right)\,\left({100\over \chi}\right)\, {\rm cm}^{-2}\,.
\end{eqnarray}
where we used realistic parametrizations for $f_e$ and $\chi$ as shown
in the subsequent paragraph. The uncertainty estimate of about one
order is a combination of the uncertainties attached to the central
parameters which is discussed in the following section.

\section{Quantitative discussion of uncertainties}

Main parameters and their uncertainties, which could be on the order
of a factor of a few, are discussed in the following paragraphs.

\subsection{Electron-to-proton luminosity ratio $f_e$}
Assuming that AGN are the sources of UHECRs, the ratio between
electron and proton luminosity, $f_e$ can be estimated empirically by
comparing the average energy density rate, $\dot{\rho}_e$ (units:
erg/(Mpc$^3\cdot$ yr)) is received from integrating over all
synchrotron output from AGN, using the RLF mentioned above;
$\dot{\rho}_{\rm CR}$ is received by integrating over the observed CR
spectrum from $E_{\rm p}^{\min}$. For $E_{\rm p}^{\min}\approx 3\cdot
10^{18}$~eV,  $f_e\approx0.01$ (FR-I) and $f_e\approx0.4$
(FR-II). Such an approach is common to use in order to correlate
possible cosmic ray sources with the observed flux of cosmic rays, see
e.g.\ \cite{waxman_bahcall1999}. While \cite{waxman_bahcall1999} apply
this strategy to Gamma-ray bursts, we use it for FR-I and FR-II
galaxies. The exact value is subject to uncertainties as discussed in
the following paragraph. It should be noted that the same procedure
can be done for starburst galaxies, and it can be shown that also these sources can provide
an energy budget which can match the IceCube observations \citep{loeb_waxman2006,becker2009}. Future measurements will be able to distinguish the two
scenarios: while an energy cutoff at PeV energies is expected in the
case of starburst galaxies, the hypothesis that the detected neutrinos
come from AGN implies that their cutoff must lie at much higher
energies, close to $10^{3}$~PeV or even larger.

This value becomes {\it smaller} the lower the minimum energy is
assumed to be. On the other hand, it {\it rises} when including
possible contributions to the electron luminosity that are not
radiated at radio wavelengths (see Section \ref{chi:sec}).

From theoretical considerations (see e.g.\ \cite{schlickeiser2002,merten2013}), for equal spectral indices of electrons and protons at injection, the ratio of the luminosities should be $f_e\approx (m_e/m_p)^{(p-1)/2}\approx 0.02$ for a primary spectral index of $p=2$. While this value is subject to change in case of spectral indices deviating from $p=2$, the ratio is certainly to be expected to be $f_e\ll 1$. Thus, the values received for FR-I or FR-II galaxies respectively, seem to be a realistic range: $0.01<f_e<0.4$.

As it is extremely difficult to pinpoint the exact value, we start by
using $f_e=10^{-1.2}$, so that a symmetric uncertainty $\Delta f_e
\approx 10^{\pm 0.8}$ is obtained. Thus, a higher value of $f_e$ would
lead to a density increase, so the density could become at maximum a
factor of $6$ higher. 

While both theoretical and
  experimental constraints bear uncertainties, they end up in
  approximately the same range and do not allow for completely
  arbitrary values: From the simple estimate in \cite{schlickeiser2002}, it is obvious
that a larger variation than a factor of $6$ is rather unlikely. This
is also supported by the indirect observational constraints from
extragalactic cosmic ray sources as discussed above. In addition, the
observation of Galactic cosmic ray sources allow for a direct
comparison of electrons to protons, as the total energy budget in
electrons can be measured directly. Here, a ratio of 1 electron for
100 protons is present. This ratio could be somewhat higher due to
significant loss processes of the electrons during propagation, mainly
through synchrotron and Inverse Compton processes, but including such
processes is not expected to contribute with a factor of $100$, so
that the statement of $f_e\ll 1$ still holds.
\subsection{Radio-electron correlation $\chi$\label{chi:sec}}

In the above calculation, it is assumed that the electron luminosity
corresponds to a factor of a few of the observed radio luminosity,
where we choose $\chi=100$, as we explain here: It is clear that synchrotron radiation from electrons is distributed over a wider energy range and that not necessarily all energy is radiated. Assuming that the relativistic electrons have a power law distributed energy with a spectral index $p$ and predominantly lose their energy via synchrotron emission, $\chi$ is subsequently determined by the ratio of electron and radio emissivity. In the case of $p=2$ the electron emissivity (in units of eV cm$^{-3}$ ster$^{-1}$ s$^{-1}$) yields $\rho_e\propto \ln(\gamma_{\rm max}/\gamma_{\rm min})$, where $\gamma_{\rm min}$ and $\gamma_{\rm max}$ is the minimal and maximal Lorentz factor of the electrons, respectively. The radio emissivity $\rho_{\rm radio}$ is determined by integrating the synchrotron emission coefficient in the radio band, i.e.\ between $\nu_{\rm min}=100\,\text{MHz}$ and $\nu_{\rm max}=5\,\text{GHz}$. Since the radio emission is determined by the rising part of the synchrotron emission spectrum, the spectral synchrotron power is accurately approximated by \cite{crusius_schlickeiser1986}, $P(\nu,\,\gamma)=1.19\, P_0\,(\nu/(\nu_s\,\gamma^2))^{1/3}\,H[\nu_s\gamma^2-\nu]$, with $P_0=2.64\times10^{-10}(B/1\,\text{G})\,\text{eV}\,\text{s}^{-1}\,\text{Hz}^{-1}$ and $\nu_s=4.2\times10^6\,(B/1\,\text{G})\,\text{Hz}$. 
Thus, the spectral cut-off by the Heaviside function yields in the case of $p=2$ and $\nu_s\gamma_{\rm max}^2>\nu_{\rm max}>\nu_{\rm min}$ the following three different $\chi$-dependencies
\begin{widetext}
\begin{equation}
\chi={\rho_e \over \rho_{\rm radio}}={32\pi\,m_ec^2\over 3\,P_0\,\tau_0}\,\nu_s^{1\over 3}\,\ln(\gamma_{\rm max}/\gamma_{\rm min})\cdot\begin{cases}(a-b)^{-1}\,,&\text{for }\nu_s\,\gamma_{\rm min}^{2}<\nu_{\rm min}\,,\\
(c_s+a_s-b_s)^{-1}\,,&\text{for }\nu_{\rm min}\leq\nu_s\,\gamma_{\rm min}^{2}\leq\nu_{\rm max}\,,\\
d^{-1}\,,&\text{for }\nu_s\,\gamma_{\rm min}^{2}>\nu_{\rm max}\,,\end{cases}
\end{equation}
with the synchrotron cooling timescale $\tau_0=7.7\cdot 10^8\,(B/(1\,{\rm G}))^{-2}\,{\rm s}$. 
\end{widetext}
The parameters are: 
\begin{eqnarray}
a=&\nu_s^{4\over 3}\ln\left(\nu_{\rm max}/\nu_{\rm min}\right)\,,\\
b=&{3\over 4}\gamma_{\rm max}^{-{8\over 3}}\left(\nu_{\rm max}^{4\over 3}-\nu_{\rm min}^{4\over 3}\right)\,,\\
d=&{3\over 4}\left(\gamma_{\rm min}^{-{8\over 3}}-\gamma_{\rm max}^{-{8\over 3}}\right)\left(\nu_{\rm max}^{4\over 3}-\nu_{\rm min}^{4\over 3}\right)\,,\\
a_s=&\nu_s^{4\over 3}\ln\left(\nu_{\rm max}/(\nu_{s}\gamma_{\rm min}^2)\right)\,,\\
b_s=&{3\over 4}\gamma_{\rm max}^{-{8\over 3}}\left(\nu_{\rm max}^{4\over 3}-(\nu_{s}\gamma_{min}^2)^{4\over 3}\right)\,,\\
d_s=&{3\over 4}\left(\gamma_{\rm min}^{-{8\over 3}}-\gamma_{\rm max}^{-{8\over 3}}\right)\left((\nu_s\gamma_{\rm min}^2)^{4\over 3}-\nu_{\rm min}^{4\over 3}\right)\,.
\end{eqnarray}
Thus, the spectral cut-off by the Heaviside function yields three
different $\chi$-dependencies at (1.) $\nu_s<\nu_{\rm
  min}\,\gamma_{\rm min}^{-2}$, (2.) $\nu_{\rm min}\,\gamma_{\rm
  min}^{-2}\leq\nu_s\leq\nu_{\rm max}\,\gamma_{\rm min}^{-2}$ and (3.)
$\nu_s>\nu_{\rm max}\,\gamma_{\rm min}^{-2}$.
Consequently, $\chi$ depends on the magnetic field strength of the considered emission area,
where $B$ generally decreases with increasing distance from the central engine of the AGN and therefore varies between some mG to a few kG.
\begin{figure}[h!]
  \centering
   \includegraphics[width=0.45\textwidth]{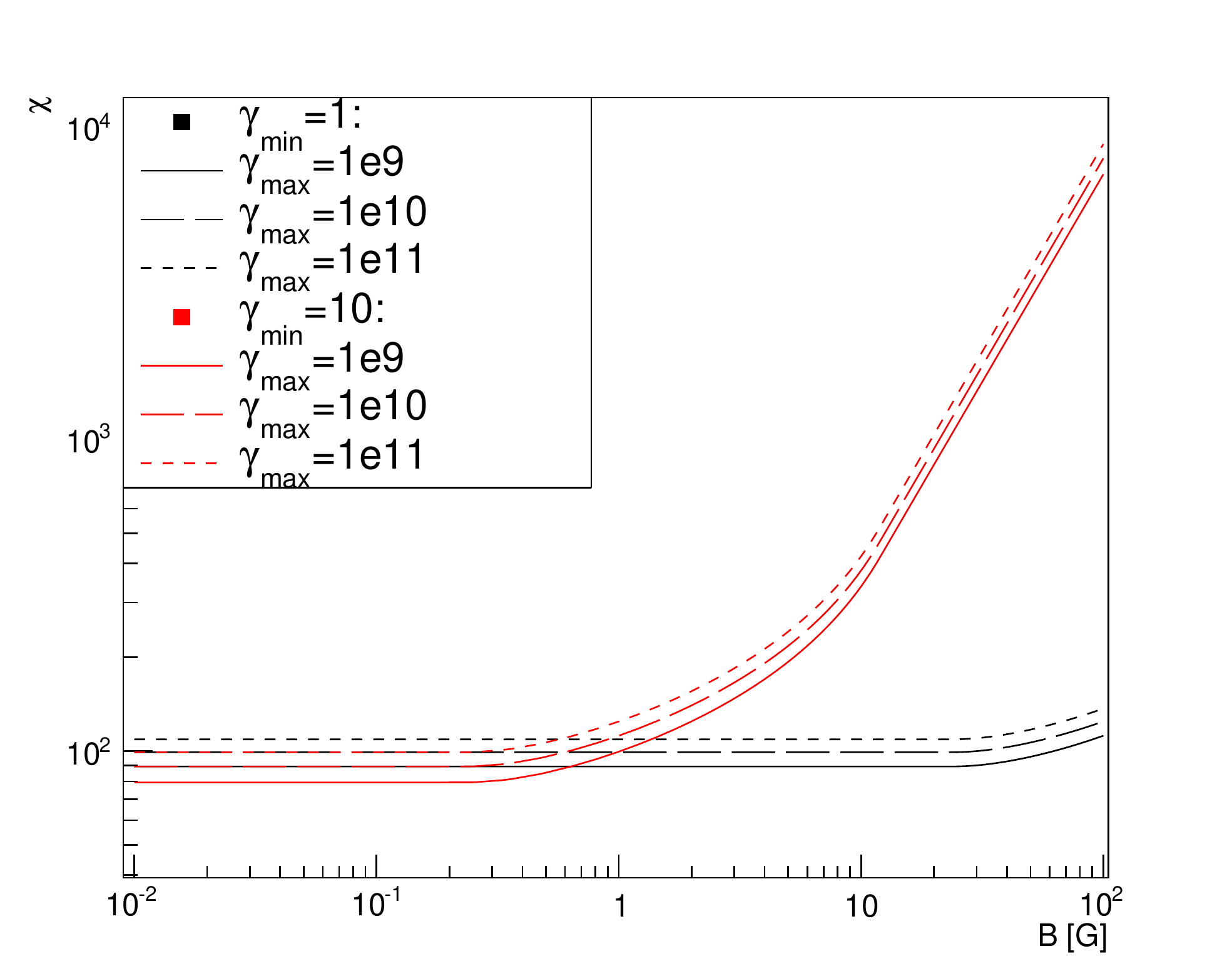}
 \caption{Dependence of the factor $\chi$ on the magnetic field strength $B$. Uncertainties from the primary electron spectrum, i.e.\ maximum and minimum Lorentz factor $\gamma$, lie below $10^{\pm0.2}$.}
\label{log_chi}
\end{figure}

 Figure \ref{log_chi} shows the dependence of $\chi$ on the magnetic field strength for different choices of $\gamma_{\min}$ and $\gamma_{\max}$.  Since FR-I and FR-II galaxies emit a significant amount of energy at radio energies, the electrons are expected to cool down till a minimal Lorentz factor $\gamma_{\rm min}\leq\sqrt{\nu_{\rm max}/\nu_s}\simeq 10^{3/2}\,(B/1\,\text{G})^{1/2}$, so a value somewhere in between $\gamma_{\min}=1-10$. The maximum energy reached in the acceleration process itself (not including losses, only acceleration) must be around $\gamma_{\max}=10^{10\pm1}$ in order to explain the observed cosmic ray spectrum which reaches up to $E_{\rm CR,max}\approx 10^{20}$~eV. This maximum energy, if dominated by iron, could be a factor of $Z = 26$ lower for protons due to the dependence of the acceleration process on the charge $Z$. Thus, a range of $\gamma_{\max}= 10^{9}-10^{11}$ seems plausible. The uncertainty of the maximal Lorentz factor of the electrons produce only an uncertainty factor of about $\Delta\chi\approx 10^{\pm 0.2}$. The choice of the minimum of the Lorentz factor determines at which critical magnetic field strength $B_{\rm c}$ the factor $\chi$ goes from being constant to increasing with a power-law (see Fig.\ \ref{log_chi}). At the most extreme case of $\gamma_{\min}=1$, $\chi$ becomes significantly larger from $B_{\rm c}\sim 10$~Gauss, for $\gamma_{\min}=10$, the relation between $\chi$ and $B$ stays approximately constant within a factor of $2$ below $B<30$~Gauss.

We thus find that $\chi$ is constant around $100$ for magnetic fields of $B<10$~Gauss and that it increases at higher magnetic fields. We take into account this behavior in the interpretation of our results.
 In general, $\chi$ is around $10^{2\pm0.2}$ and independent of the magnetic field strength when the emission region is at a distance of more than about a parsec from the central engine of the AGN due to the correlated $B$-regime where $\nu_s\,\gamma_{\rm min}^{2}<\nu_{\rm min}$.
\subsection{Radio Luminosity Function}
The two AGN populations used here, FR-I and FR-II galaxies, represent the two most extreme scenarios of source evolution, one population having a large contribution from low-luminosity sources, one being focused on high-luminosity sources. The final result is still somehow compatible, as the differences in redshift dependence and luminosity dependence cancel out. If one separately considers the differences in the results for $\zeta_L$ and $\zeta_z$, there is a factor of $\sim 10$ variation in each of the factors. When comparing the same source classes, the uncertainties are expected to be much smaller, on the order of a factor of $\sim 2-3$ for the product of $\zeta_L$ and $\zeta_z$. The main reason is that both factors are mainly dominated by the integration limits, as they have very strong evolving integrands. So, changing the functions themselves does not change too much in the total result. We thus apply a maximum of a factor of $3$ uncertainty from this, so $10^{\eta\pm 0.5}$, where $\eta=\log[(\zeta_z\cdot \zeta_L)/(\text{GeV}\,\text{Mpc}^{-2}\,\text{s}^{-1}\,\text{sr}^{-1})]$ for FR-I and FR-II galaxies.

%%%%%%%%%%%%%%%%%%%%%%%%%%%%%%%%%%%%%%%%
\section{Discussion of the results}
%%%%%%%%%%%%%%%%%%%%%%%%%%%%%%%%%%%%%%%%
In the previous sections, we showed that proton-proton interactions
can produce a neutrino signal of a given strength for a fixed
combination of magnetic field strength $B$ and column depth
$N_H=n_H\cdot R$ at the source. Uncertainties in the calculation
of approximately one order of magnitude are applied using an uncorrelated
Gaussian error estimate to combine the uncertainties in the parameters
discussed above. This constrains the possible
acceleration site in the $(B,N_H)$-space. 

\begin{figure}[h!]
  \centering
    \includegraphics[width=0.5\textwidth]{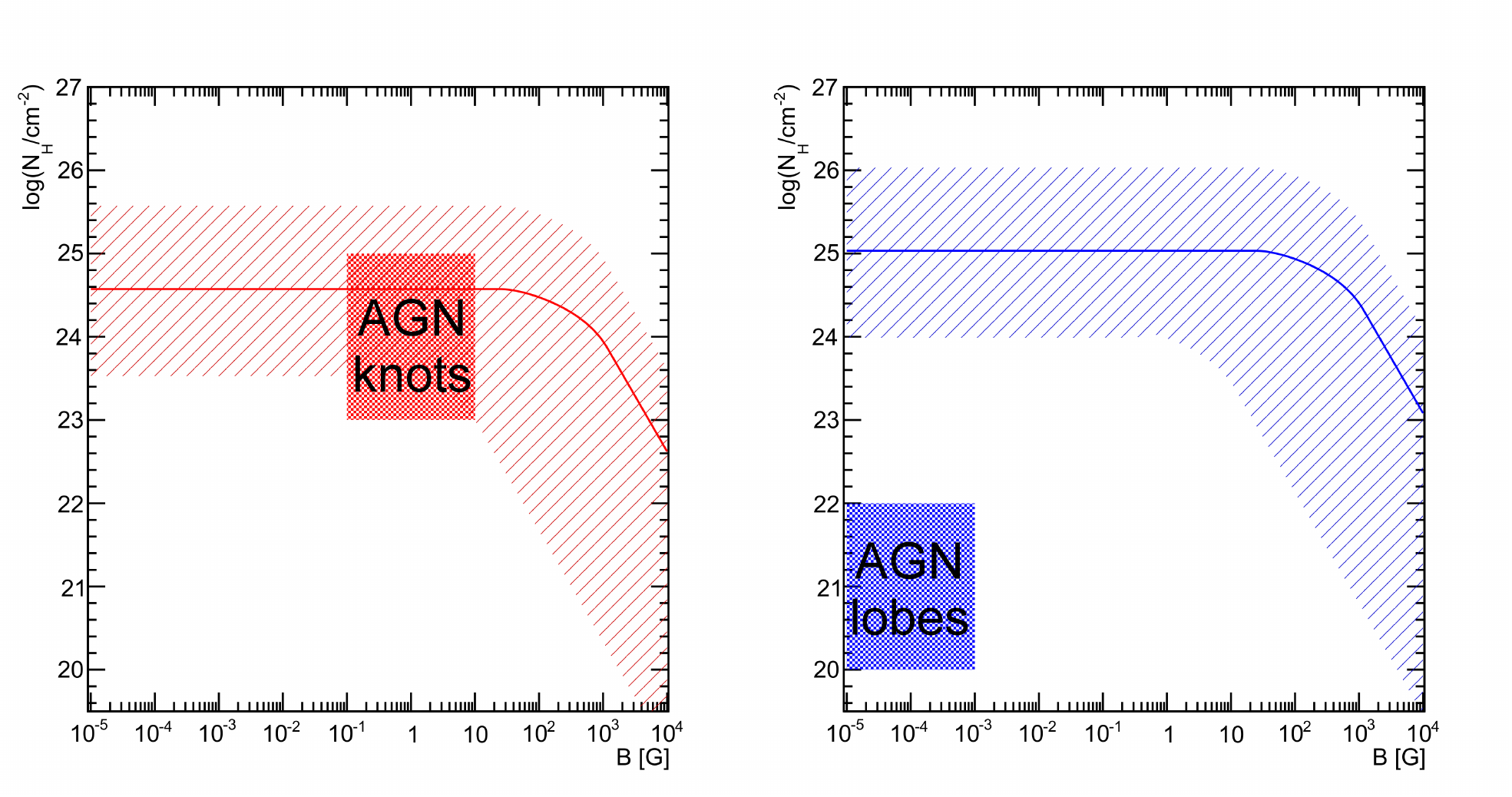}
  \caption{Allowed parameter range for column density $N_H$ and magnetic field strength $B$ in FR-I (left panel) and FR-II (right panel) galaxies. The dashed areas represent the regions derived including uncertainties in the calculation, dominated by the parameters $\eta$, $\chi$ and $f_e$, as discussed in the text. 
The encircled areas mark the approximate position of the knots and lobes, producing the radio signal in the respective calculation.}
\label{figfr1-2}
\end{figure}

The results are shown in Fig.\ \ref{figfr1-2}. 
The shaded band represents the parameter space for $(N_H,\,B)$ derived from the IceCube observations, applying the above-discussed error of $10^{\pm1.0}$ to the region in which the parameter $\chi$ is constant, i.e.\ for $B<B_c$ as discussed before. At higher magnetic fields, we show the range possible for $1<\gamma_{\min}<10$.

The radio emission from electron synchrotron radiation, 
used to determine the neutrino flux, comes from the knots in the case
of FR-I galaxies and from the lobes for FR-II galaxies. We therefore
compare the shaded band for FR-I galaxies with the approximate
parameters in the knots. For the calculation of the column depth, we
assume a density of $\sim 10^{9}$~cm$^{-3}$ and a knot size of
$10^{-3}$~pc close to the foot of the jet, see
e.g.\ \cite{eichmann2012}. As the density decreases, the knot size
increases with the distance $z$ from the foot of the jet, so that the
column is expected to stay approximately the same. The most important
contribution is expected to come from the foot of the jet (see
e.g.\ \cite{becker_biermann2009}, so that this assumption is
reasonable. The magnetic field decreases with the distance along the
jet $z$ as well, $B(z)\sim B_0\cdot (z/z_0)^{-1}$
\cite{osullivan2009}, see \cite{becker_biermann2009} for a discussion
of neutrino production in that context. In the graph, we indicate the
highest magnetic fields, $B_0\sim 0.1-10$~Gauss. For lower fields,
which should be present along the jets for large $z$
\cite{kataoka_stawarz2005}, our results do not change. These
considerations result in a possible parameter range for FR-I galaxies
of $(N_H,\,B)=(10^{24\pm1}$cm$^2,\,10^{0.5\pm0.5}$~Gauss$)$. This
realistic range of parameters for FR-I galaxies is now compared with
the allowed range if the IceCube signal should be explained by
emission from FR-I galaxies. This is shown in the left panel of
Fig.\ \ref{figfr1-2}. The knots fall right into the allowed region and
we therefore consider FR-I galaxies as a serious candidate as the
sources for the detected IceCube signal. 
The right plot of Figure \ref{figfr1-2}, on the other hand, shows that
FR-II lobes are far too less dense to produce the signal. For the
calculation of the column depth present in FR-II radio lobes, we
assume a density of $0.01-0.1$~cm$^{-3}$, as the jets meet the
intergalactic medium, and a lobe size of $10^{22}-10^{23}$~cm, see
e.g.\ \cite{wellmann1997}. The approximate value of the magnetic field
is taken from \cite{kataoka_stawarz2005},
i.e.\ $(N_H,\,B)=(10^{21\pm1}$cm$^2,\,10^{-4\pm1}$~Gauss$)$ for the
lobes. Thus, proton-proton interaction in radio lobes of FR-II galaxies can be excluded as the
sources of the IceCube signal.
 It should be
noted that this discussion only includes proton-proton interactions, and
does not take into account photohadronic interactions of cosmic rays
with ambient photon fields. In principle, proton-photon interactions could contribute to a possible signal in the lobes (see also
\cite{winter2013} for a discussion). As it is kinetically necessary
to produce the delta resonance, however, a relatively high-energy
photon field needs to be present in order to produce a high optical
depth for the process. With the dominant electromagnetic emission
coming from radio wavelengths in the lobes, this seems rather
unlikely. 

\begin{figure}[h!]
  \centering
    \includegraphics[width=0.5\textwidth]{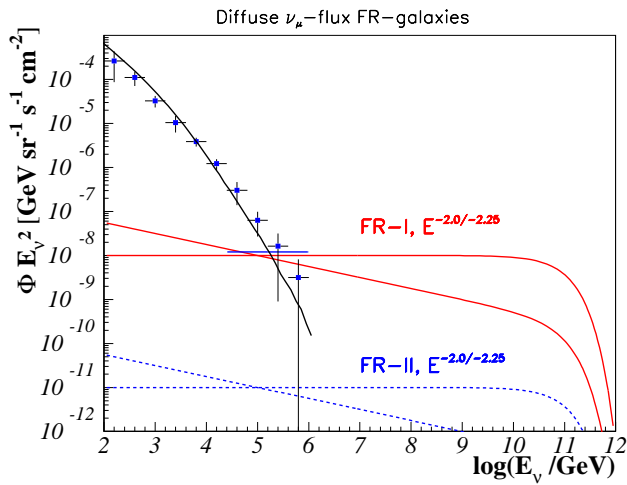}
  \caption{Expected neutrino flux for FR-I galaxies (solid, red
    line) and FR-II galaxies (dashed, blue line). For FR-I galaxies, an average value for the column depth of
    $N_H=10^{24.5}$~cm$^{-2}$ and a magnetic field $B<10$~Gauss are
    used, which are realistic parameters (see
    Fig.\ \ref{figfr1-2}). In addition, we show the potential flux,
    close to what is expected from IceCube if the measured flux is
    steeper than $E^{-2}$. Concretely, we show a proton spectrum of
    $E^{-2.2}$, which translates to a neutrino spectrum close to
    $E^{-2.25}$. For FR-II galaxies, we use the most
    optimistic assumption of a column depth of $10^{22}$~cm$^{-2}$ and
    a B-field of $B<10$~Gauss. The atmospheric measurement is taken
    from \cite{tim_icrc2013} and the atmospheric prediction represents
    the model of \cite{anatoli2012}. For a
  colored version of the graph, see online version of the paper. }
\label{flux:fig}
\end{figure}

In order to show what our results mean in terms of the absolute
neutrino flux, we show the estimates for FR-I and FR-II galaxies in
Fig.\  \ref{flux:fig}. For FR-I galaxies in the case of an $E^{-2}$ spectrum, we use a column of $N_H\sim
10^{24.5}$ and assume that the magnetic field on average is
lower than $10$~Gauss. We also show a spectrum corresponding to an
$E^{-2.2}$ proton injection spectrum. Here, we
use a column of $10^{23.6}$, required to approximately match the IceCube
data. This cannot be done in an exact way, as firstly IceCube only provides
values for an $E^{-2}$ spectrum and secondly, the dependence on the
exact choice of the minimum energy becomes relevant, which we chose to
be $E_{\min}=100$~GeV here. In this approximate way, the number is
compatible with what is expected from the observation of the column
from radio galaxies. The general result does not change for an $E^{-2.2}$ spectrum: FR-I galaxies are still well-compatible with the observations, while FR-II galaxies have too low columns.

The flux is well-suited to explain the IceCube
signal. It should be noted here that more tests are clearly necessary
to prove (or disprove) this model, as the attached uncertainty is
still about one order of magnitude as discussed above. A smoking gun
would of course be the detection of the nearest point sources, which
would be M87 and CenA \cite{saba2013}, or possibly a stacked signal of
the nearest FR-I galaxies, see \cite{stacking_interpret2007} and references therein. Further, future observations by
IceCube will show if the spectrum really does persist beyond PeV
energies or if there is a cutoff at PeV energies. In the latter case,
AGN can be excluded if the flux should at the same time be associated
with the production of UHECRs. In that case, a cutoff in the spectrum
should only be present at $\sim 10^{3}-10^{4}$~PeV. On the other hand, AGN
models are very well compatible with energy spectra slightly steeper
than $E^{-2}$. Gamma-ray observations of CenA and M87 would even
indicate a spectral behavior close to $E^{-2.3}$ rather than $E^{-2}$
(see \cite{saba2013} and references therein). 

Compatibility difficulties with the extragalactic gamma-ray background as measured by Fermi are discussed 
in \cite{murase2013}. The comparison of the photon background measured with Fermi
and the neutrino signal has to deal with different uncertainties: first of all, the total luminosities need to be compared,
as at least a fraction of the gamma-rays cascades down from higher energies by interactions with the extragalactic background light. For the
luminosities of both the neutrinos and photons, extrapolations into an unknown parameter space is necessary. For neutrinos, in particular the lower
integration threshold is relevant, for Fermi, the higher integration threshold is not exactly known. In addition, the possible contribution from sources at
low redshift leaves the possibility of a larger fraction of TeV-PeV sources to contribute significantly at a level that is not known at this point. In addition, 
the effects of uncertainties in the luminosity dependence of the source evolution function is another source of uncertainty. We therefore consider these results
for an $E^{-2.3}$ spectrum compatible with Fermi and IceCube data. 
If a such a spectral
behavior would be confirmed by IceCube, it would be fully compatible
with the model of neutrino emission from AGN. 

 For FR-II galaxies, we use the most optimistic
case of a column depth of $N_{H}\sim 10^{22}$~cm$^{-2}$. It would be
extremely difficult to raise the level of this flux by tuning the
parameters by the three orders of magnitude needed to explain the
IceCube signal. It is obvious
from Fig.\ \ref{flux:fig} that this emission scenario can be excluded
from the possible list of sources for the IceCube signal. This result supports the study of proton-proton interactions in the lobes of Centaurus A, which are also discussed to be too weak to contribute significantly to a neutrino signal \cite{fraija2012}.

%%%%%%%%%%%%%%%%%%%%%%%%%%%%%%%%%%%%%%%%
\section{Conclusions \& Outlook}
%%%%%%%%%%%%%%%%%%%%%%%%%%%%%%%%%%%%%%%%

In this paper, we show what conditions need to prevail in an
acceleration environment in FR-I and FR-II radio jets in order to
provide a cosmic ray interaction site which is capable of explaining
the observed IceCube signal. We assume that leptonic and hadronic
cosmic rays are accelerated at the same site at a constant luminosity ratio and that the observed
synchrotron radiation from AGN represents a part of the energy budget
available in cosmic ray electrons. The exact fraction of
radio-to-electron energy depends on the magnetic field at the
acceleration site, which turns out to be one of the free parameters
connected to the acceleration site. A second parameter in the
calculation is the column depth at the interaction site. 

We estimate the uncertainties connected to the determination of the column depth in dependence on the magnetic field. For the electron-to-proton ratio, we argue that this lies at $f_e=10^{-1.2\pm 0.8}$. The factor $\chi$ is shown to be known within $\chi=10^{2\pm0.2}$. For the luminosity and redshift factors, we have taken into account an uncertainty of $\Delta(\zeta_L\cdot \zeta_z)=10^{\pm0.5}$, associated with the uncertainty in the luminosity function.

Considering
the observed flux of high-energy neutrinos with IceCube at a level of
$10^{-8}\,\frac{\rm GeV}{{\rm cm}^2\,{\rm s}\,{\rm sr}}\cdot
E_{\nu,0}^{-2}$, we find that for magnetic fields at the acceleration
site of $B<10$~Gauss, a column depth of $N_H\sim 10^{24.5}$~cm$^{-2}$ (FR-I) and $N_H\sim 10^{25}$~cm$^{-2}$~cm$^{-2}$ (FR-II) 
is needed in order to explain the observed astrophysical signal as
coming from FR-I or FR-II radio jets, respectively. For higher magnetic fields, the
column depth must be lower. This is an effect of decreasing
contribution of the electron population to the flux radiated at radio
wavelengths. Here, we discuss two scenarios as examples:
\begin{enumerate}
\item Acceleration and interaction in AGN knots: with a column of
  $\sim 10^{24\pm1}$~cm$^{-2}$ and a magnetic field of around $1-10$~Gauss we
  find that AGN knots are well-suited to explain the observed signal
  with proton-proton interactions from FR-I galaxies.
\item Acceleration and interaction in AGN lobes of FR-II galaxies: here, the column
  depth is too low $\sim 10^{21\pm1}$~cm$^{-2}$ at a given magnetic field
  of $\sim 10^{-4\pm1}$~Gauss in order to explain the signal with
  proton-proton interactions. It might still be possible to produce the
  neutrino flux via photohadronic interactions.
\end{enumerate}

In the future, IceCube will be able to provide valuable results concerning the spectral behavior of the energy spectrum and the direction of the events. In the near future, the more detailed determination of the spectral behavior will already help to further exclude source models. The model presented here predicts that a spectrum of around $E^{-2.3}$ persists up to far beyond PeV energies. This condition comes from the assumption that these neutrinos are directly connected to the extragalactic flux of ultra-high energy cosmic rays. If a cutoff at PeV energies is observed, the sources proposed here can be excluded as a possible class for the detected neutrinos. In that case, starburst galaxies, with a cutoff below or probably at 1 PeV would be an interesting alternative, see e.g.\ \cite{loeb_waxman2006,becker2009}.

In the more distant future, once the exact point sources responsible for the so-far diffuse high-energy neutrino signal can be identified, it will be easier to investigate both the source class and the exact emission region within the specific source and by that identify the sources of UHECRs. The relation between the diffuse neutrino flux and the contribution from point sources will provide information on the luminosity function of the sources of UHECRs. Another important piece of information will be provided through the exact measurement of the spectral behavior of the astrophysical flux. At this point, the energy distribution indicates that there is either a cutoff present at PeV energies, or that the spectrum is somewhat steeper than $E^{-2}$, i.e.\ $\sim E^{-2.2}$. While an AGN hypothesis is well-compatible with an $E^{-2.2}-$spectrum, a cutoff at PeV-energies would at least not be compatible with the hypothesis of a connection to the observed cosmic ray flux above the ankle. Future measurements will be able to resolve this question.

\noindent

\acknowledgements{We thank Peter L.\ Biermann, Thorsten Gl\"usenkamp,
  Alexander Kappes, Nils Nierstenh\"ofer, Wolfgang Rhode, Reinhard Schlickeiser, Sebastian Sch\"oneberg and the
   IceCube Collaboration for very helpful discussions. Thank you also to the useful comments from the anonymous referees. JT, SMS and BE
 acknowledge support from the Deutsche Forschungsgemeinschaft (DFG),
 grant BE3714/4-1 and the MERCUR project Pr-2012-0008, as well as from the Research Department of Plasmas with Complex Interactions (Bochum).}

\end{document}